\documentclass[aip,apl,reprint,numerical]{revtex4-1}

\usepackage[T1]{fontenc}
\usepackage[utf8]{inputenc}

\usepackage[version=4]{mhchem}
\usepackage{todonotes}
\usepackage{graphicx}
\usepackage{siunitx}
\usepackage{amsmath}
\usepackage{amssymb}
\usepackage{float}

\usepackage{comment}

\draft 

\begin{document}


\title{Contact-free reversible switching of improper ferroelectric domains by electron and ion irradiation} 



\author{Erik D. Roede}
\affiliation{Department of Materials Science and Engineering, Norwegian University of Science and Technology (NTNU), 7491 Trondheim, Norway}

\author{Aleksander B. Mosberg}
\affiliation{Department of Physics, Norwegian University of Science and Technology (NTNU), 7491 Trondheim, Norway}
\author{Donald M. Evans}
\affiliation{Department of Materials Science and Engineering, Norwegian University of Science and Technology (NTNU), 7491 Trondheim, Norway}
\author{Edith Bourret}
\affiliation{Materials Sciences Division, Lawrence Berkeley National Laboratory, Berkeley, California 94720, USA}
\author{Zewu Yan}
\affiliation{Department of Physics, ETH Zurich, Otto-Stern-Weg 1, 8093 Zurich, Switzerland}
\author{Antonius T. J. van Helvoort}
\affiliation{Department of Physics, Norwegian University of Science and Technology (NTNU), 7491 Trondheim, Norway}
\author{Dennis Meier}
\email[]{dennis.meier@ntnu.no}
\affiliation{Department of Materials Science and Engineering, Norwegian University of Science and Technology (NTNU), 7491 Trondheim, Norway}

\date{\today}

\begin{abstract}
Focused ion beam (FIB) and scanning electron microscopy (SEM) are used to reversibly switch improper ferroelectric domains in the hexagonal manganite \ce{ErMnO3}. Surface charging is achieved by local ion (positive charging) and electron (positive and negative charging) irradiation, which allows controlled polarization switching without the need for electrical contacts. Polarization cycling reveals that the domain walls tend to return to the equilibrium configuration obtained in the as-grown state. The electric field response of sub-surface domains is studied by FIB cross-sectioning, revealing the 3D switching behavior. The results clarify how the polarization reversal in hexagonal manganites progresses at the level of domains, resolving  both domain wall movements and the nucleation and growth of new domains. Our FIB-SEM based switching approach is applicable to all ferroelectrics where a sufficiently large electric field can be built up via surface charging, facilitating contact-free high-resolution studies of the domain and domain wall response to electric fields in 3D.

\end{abstract}


\maketitle 

\section{Introduction}
Ferroelectric switching is at the heart of key emerging technologies, including ferroelectric memory\cite{Arimoto2004,Garcia2014}, transistors\cite{Si2019} and ferroelectric catalysts\cite{Kakekhani2015}. In these applications, the ability to set the ferroelectric polarization into two or more stable states is utilized to store information, gate electrical currents and control photocatalytic properties, respectively. At its roots, polarization switching in ferroelectrics corresponds to an interaction of its electric domains. In order to improve performance and achieve active devices with new functional properties, it is crucial to understand how the electric polarization transitions between  different energy-equivalent states. For proper ferroelectrics, where the spontaneous polarization is the symmetry breaking order parameter, electric switching has been studied extensively and comprehensive theories that describe the domain reversal in applied electric fields are established \cite{Tagantsev2010}.

Recently, materials where an electrical polarization arises as a secondary effect, so-called improper ferroelectrics, are attracting increasing attention. In these materials, ferroelectricity is induced, e.g., by a lattice distortion or magnetic order, giving rise to additional functional properties beyond just ferroelectricity\cite{Evans2020, Meier2020}. Examples of such additional degrees of freedom include the unusual multiferroic hybrid domains that form in spin-driven ferroelectrics\cite{Meier2009,Meier2009a} and the stabilization of charged domain walls with unique electronic transport properties in systems with geometrically driven ferroelectricity\cite{Meier2012, Wu2012}. Despite the intriguing functional properties of improper ferroelectrics and the growing interest in this class of materials, their complex switching behavior at the level of domains and domain walls are just now beginning to be understood\cite{Huang2016, Matsubara2015, McQuaid2017}. One of the biggest challenges lies in the realization of minimally invasive experiments that allow for controlling and resolving the improper ferroelectric domains without affecting or covering their intrinsic response.

The family of hexagonal manganites (\ce{$R$MnO3}, \ce{$R$ = Sc}, Y, In, Dy$-$Lu) is a model system for improper ferroelectricity, where the electric polarization ($P_s = \SI{5.6}{\micro\coulomb\per\square\centi\metre}$) \cite{Coeure} emerges as a byproduct of a lattice-trimerizing structural distortion\cite{Meier2013, Aken2004, Fennie2005}. Building on first hysteretic switching experiments in 1963 \cite{bertaut1963} and optical imaging of the characteristic six-fold domain structure in 1967 \cite{safrankova1967}, more recent atomic force microscopy (AFM) experiments revealed an unusual electric-field response at the domain level \cite{Choi2010,Jungk2010}. It was observed that when scanning with an electrically biased AFM probe tip, the energetically unfavorable polarization domains shrink. However, in contrast to conventional ferroelectrics, topological protection prevents the system from reaching a mono-domain state\cite{Choi2010}. Indeed, calculations by Yang et al. \cite{Yang2017} predict that the topologically protected domains are stable up to an electric field strength of about \SI{241}{\kilo\volt\per\centi\metre}.

The unusual behavior observed in spatially resolved measurements motivated broader research activities, revisiting the switching behavior in hexagonal manganites across all relevant length scales from atomic to macroscopic distances. At the atomic scale, \emph{in-situ} electric field poling experiments in scanning transmission electron microscopy (STEM) \cite{Han2013} revealed that domains can shrink down to the dimension of unit cells without vanishing. Despite the topological protection, the macroscopic response is very similar to conventional proper ferroelectrics \cite{Ruff2018}, which has recently also been observed at the domain level by AFM-based switching experiments\cite{Kuerten2020}. Due to the formation of a barrier layer at the electrode-sample interface,  ferroelectric switching was only achieved at low temperature $(T\lessapprox \SI{160}{\kelvin})$ \cite{Kuerten2020,Ruff2018}. 

At low temperatures, however, charge carriers within the material (electronic and/or ionic) are effectively immobile and, hence, not readily available to screen emergent surface and domain-wall bound charges\cite{Schoenherr2019}, which can drastically alter the switching behavior. In addition, it remains unclear whether or not the observed behavior at the domain level is specific to AFM-based switching experiments and how the domain reversal observed at the sample surface progresses into the bulk.

\section{Results}
In order to facilitate ferroelectric domain switching in a contact-free fashion and study the electric field response in 3D, we used FIB-SEM. Samples with different orientation of the spontaneous polarization (in-plane (110)-oriented and out-of-plane (001)-oriented) are extracted from an \ce{ErMnO3} single crystal\cite{Yan2015} using previously established FIB lift-out protocols \cite{Mosberg2019}.

 \begin{figure}[htp]
\includegraphics{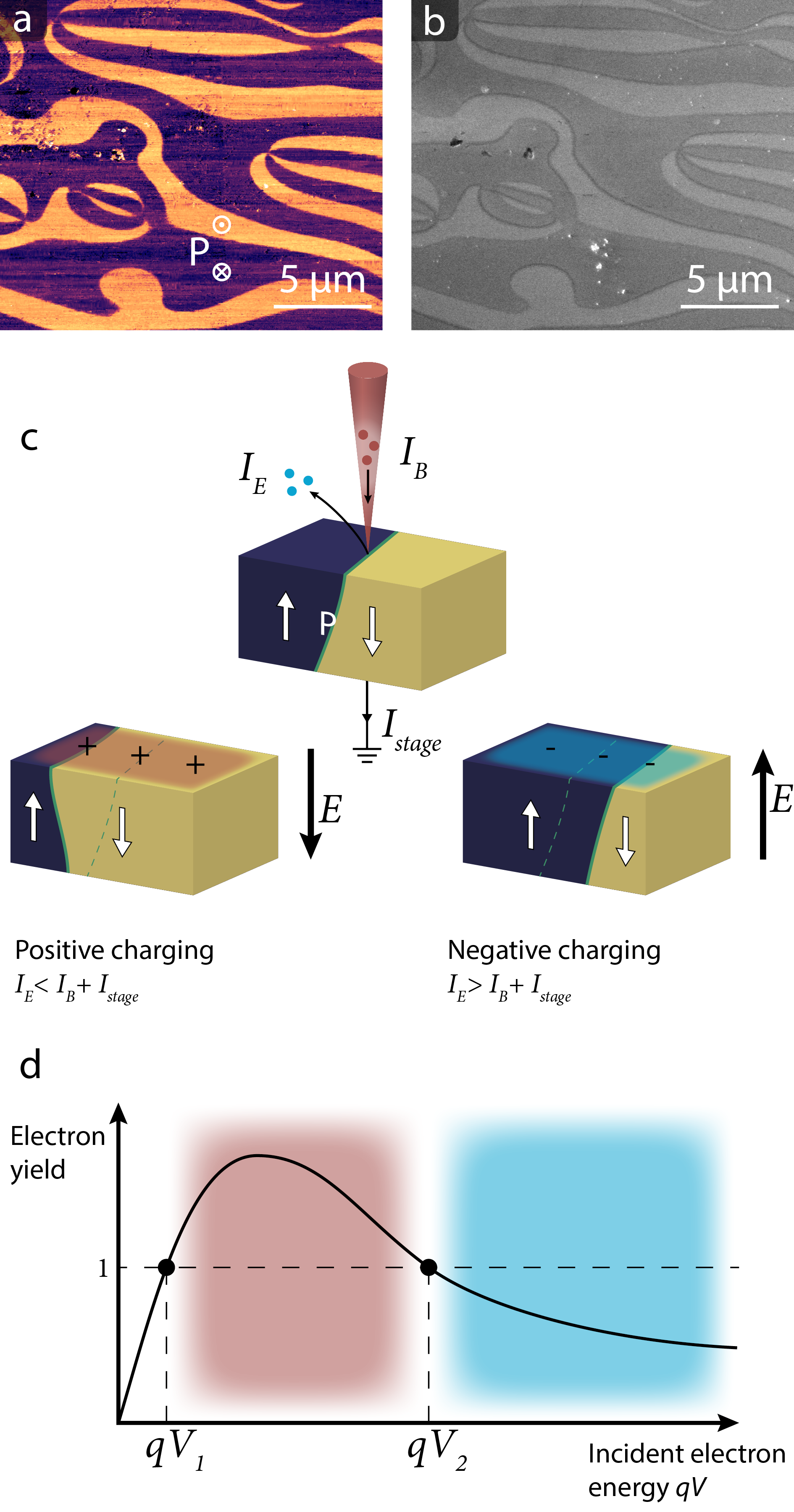}
 \caption{\label{fig0} Contrast correlation and charging mechanism. (a) PFM  and (b) SEM images of the polar surface of \ce{ErMnO3} (out-of-plane polarization). Bright SEM contrast corresponds to domains with the polarization pointing out of the surface plane. (c) Schematic illustration showing how charged particle irradiation can lead to surface charging that switches the ferroelectric polarization. $I_B$ is the incident beam current (electron or ion), $I_E$ is the current of emitted particles, $I_{stage}$ is the leakage current through the sample to the sample stage. When the currents in and out of the sample are not equal, charge accumulates. (d) General secondary electron yield as function of incident electron energy $qV$ (emitted electrons per incident electron) \cite{Reimer2010}. For acceleration voltages between $V_1$ and $V_2$ (marked in red), the sample charges positively as more electrons are emitted from the sample than impinge on it. For voltages above $V_2$ (marked in blue), the sample charges negatively. Irradiation with electrons in SEM can therefore charge a sample both positively and negatively.}
 \end{figure}
 
Figure \ref{fig0} introduces the general approach for domain imaging and switching applied in this study. Figure \ref{fig0} (a,b) shows the characteristic ferroelectric domain structure of \ce{ErMnO3} obtained on a polar surface (out-of-plane polarization) using (a) piezoresponse force microscopy (PFM) (dual-AC resonance tracking PFM at \SI{5}{\volt} peak-to-peak) and (b) SEM (\SI{3}{\kilo\volt} acceleration voltage, \SI{0.1}{\nano\ampere} nominal beam current, in-lens secondary electron detector), respectively. Comparison of the two images demonstrates that both PFM and SEM are sensitive to the ferroelectric domain distribution in \ce{ErMnO3}\cite{Li2012,Cheng2015,Rayapati2020}. In contrast to PFM, where image formation relies on differences in the electromechanical response of $\pm$P domains, SEM exploits differences in electron emission yield, allowing high-resolution microscopy experiments without the need for electrical contacts. The imaging rate of SEM is also higher than SPM, which allows capturing the dynamics with better time resolution. Under the imaging conditions applied in Figure \ref{fig0} (b), we find that $+$P domains (bright) have a larger secondary electron yield than $-$P domain (dark). This correlation between SEM contrast and polarization direction holds if imaging parameters are kept constant and can therefore be used to track changes in the domain structure. See, e.g., ref. \cite{Hunnestad2020} for a more detailed discussion of SEM domain and domain wall contrast in ferroelectrics. 

Our strategy for generating the positive and negative electric fields required for controlling the improper ferroelectric domains in \ce{ErMnO3} is presented in Figure \ref{fig0} (c), showing schematically how irradiation with charged particles can lead to either negative (electron) or positive (electron and \ce{Ga+}) surface charging. Charging in SEM is conventionally explained by the electron yield as a function of incident electron energy $qV$, where $V$ is the acceleration voltage. In general, two characteristic voltages $V_1$ and $V_2$ exist where the beam current $I_B$ is balanced by the emitted electron current $I_E$ and the leakage current $I_{stage}$.  In our experiments, stable SEM imaging with minimal charging artifacts is achieved around \SI{3}{\kilo\volt}, indicating this is close to $V_2$. For lower and higher acceleration voltages, the surface charges positively and negatively, respectively, leading to pronounced electric fields that can be used for controlling ferroelectric domains. While domain switching in ferroelectrics has been demonstrated separately using FIB, SEM and TEM \cite{Vlasov2018,McGilly2017, Chen2016, Hart2016}, the combination of FIB and SEM enables {\em{in-situ}} investigations of the domain response in 3D, as demonstrated in the following.

\begin{figure}
\includegraphics{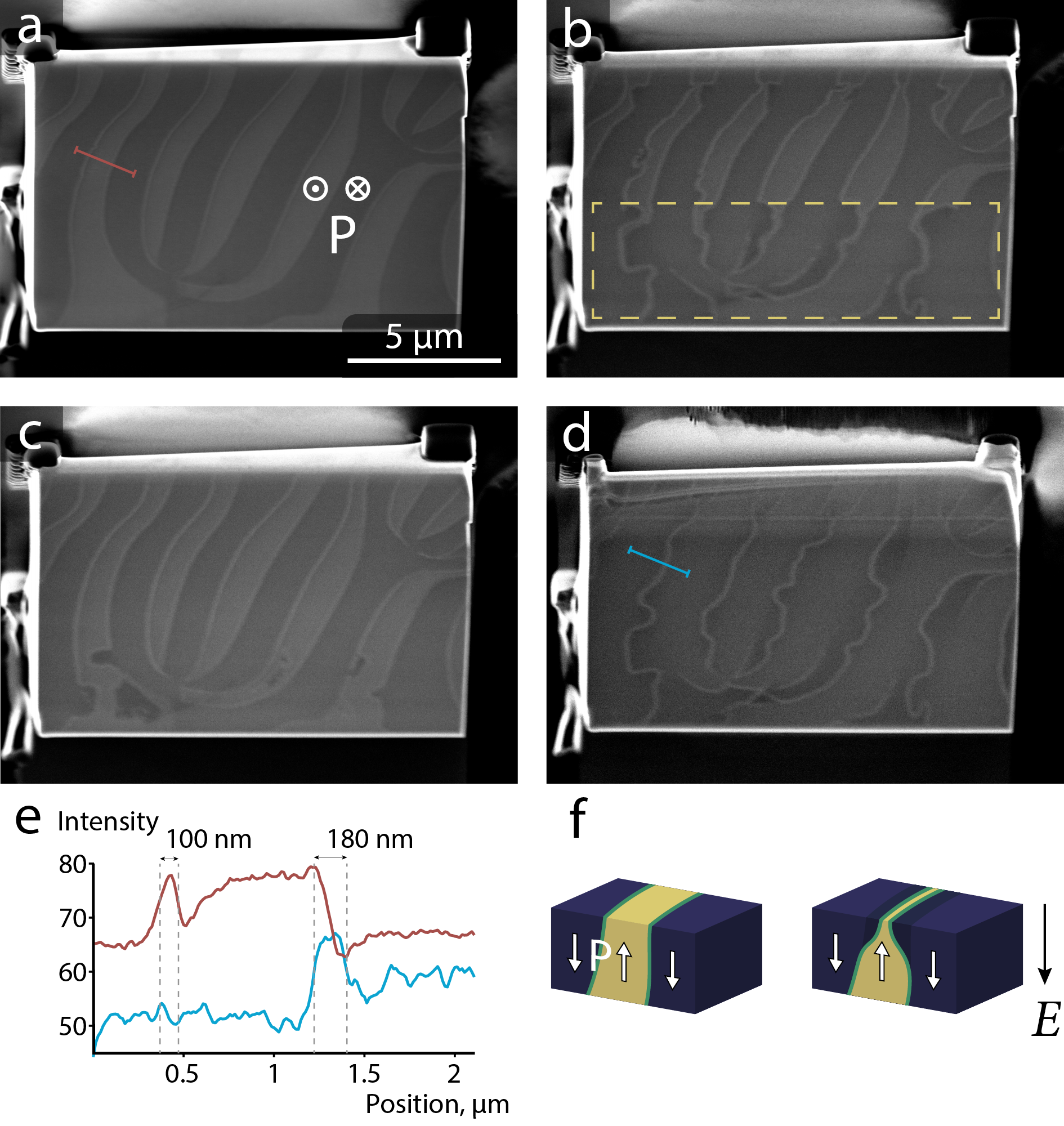}%
    \caption{\label{fig2} Reversible switching of an \ce{ErMnO3} lamella with out-of-plane polarization.(a) Domain configuration observed by SEM after lift-out. (b) After ion beam irradiation of the lower half of the sample (indicated by the yellow dashed rectangle), the $+$P domains in this region have contracted, being now visible as narrow bright bands. The original domain structure is visible as a dark contrast. (c) Electron beam exposure switches the domains back to the as-grown state, except for some areas in the lower part of the lamella. (d) Domain state obtained after exposure of the entire surface to the ion beam. (e) Intensity profile from the lines marked in (a) and (d) averaged over 15 pixels width. The width of the contracted domains is approximately \SI{180}{\nano\metre}, and the SEM signature of one domain wall around \SI{100}{\nano\metre}. (f) Proposed domain morphology upon switching with domain walls moving only near the sample surface. All SEM images taken at \SI{3}{\kilo\volt}. 
}%
\end{figure}

To analyse the response of ferroelectric domains to electron and ion irradiation in a specimen with out-of-plane polarization, Figure \ref{fig2} (a-d) shows an SEM image series recorded on a plan-view lamella (thickness $\approx$ \SI{1}{\micro\metre}) welded to a gold-coated \ce{Si} substrate. Figure \ref{fig2} (a) presents the initial ferroelectric domain structure obtained under the same imaging conditions as in Figure \ref{fig0} (b). Figure \ref{fig2} (b) shows the same area after exposing the lower part (highlighted by the yellow dashed rectangle) of the sample surface to the \SI{30}{\kilo\volt} \ce{Ga+} ion beam. The $+$P domains in the exposed area have contracted to meandering bands (we note that the original domain structure appears to be still visible, retraced by a rather faint dark contrast, which will be discussed later on). A recovery of the original domain structure is achieved by repeated imaging at \SI{3}{\kilo\volt}, leading to the domain structure in Figure \ref{fig2} (c). A comparison of Figures \ref{fig2} (a) and (c) indicates that the domain walls have returned to their original positions, with only a few exceptions in the lower part of the image. The observed domain wall pinning in this area suggests the presence of defects, causing domain wall roughening \cite{Paruch2013, Smabraten2020}. A likely source is implanted \ce{Ga+} ions or structural defects originating from the ion irradiation. Independent of the observed local pinning, however, repeated exposure to the \SI{30}{\kilo\volt} \ce{Ga+} ion beam allows driving the system back into the same poled state observed initially, as can be seen in Figure \ref{fig2} (d), showing the domain structure after exposure of the entire lamella to the ion beam. This domain poling behavior, here driven by  \ce{Ga+} exposure, is consistent with previous SPM-based poling experiments\cite{Jungk2010,Choi2010}.

\begin{figure*}
\includegraphics{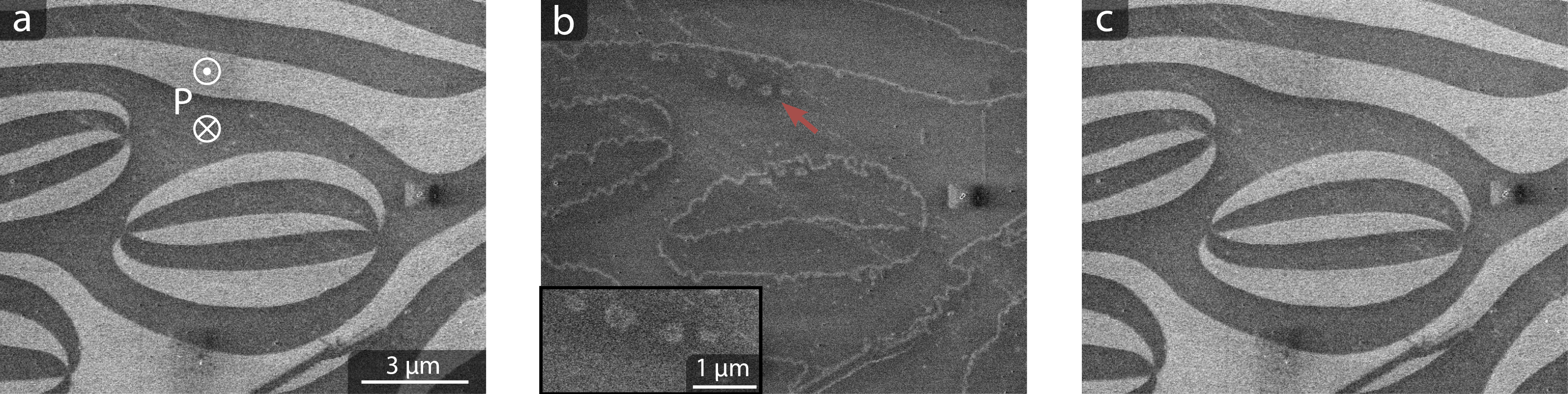}%
\caption{\label{fig3} Electron-beam induced switching on the polar surface of an \ce{ErMnO3} single crystal. Images are taken at  $\SI{1}{\kilo\volt}$, inducing positive surface charging. (a) Initial scan, (b) second scan, showing the polarization has been switched toward pointing down. In addition to the meandering line-shaped domains, loop-shaped domains are visible and marked by the red arrow. Inset: zoom-in on the loop-shaped domains.  (c) SEM scan taken after irradiating the poled area seen in (b) at  $\SI{3}{\kilo\volt}$, revealing a pattern identical to the initial domain structure. All images are taken with identical conditions (acceleration voltage: \SI{1}{\kilo\volt}, beam current: \SI{0.1}{\nano\ampere}, dwell time: \SI{5}{\micro\second}, in-lens secondary electron detector) delivering a dose of \SI{274}{\micro\coulomb\per\square\centi\metre} per scan.}
\end{figure*}

Intensity profiles (averaged over 15 pixels width) taken along the red and blue lines marked in \ref{fig2} (a) and (d) are displayed in Figure \ref{fig2} (e), showing that the $+$P domains contract to a width of approximately \SI{180}{\nano\metre}. This is slightly larger than in previous TEM- and AFM-based investigations, which reported a domain width of \SI[separate-uncertainty=true]{60(10)}{\nano\metre} \cite{Jungk2010} for electrically poled hexagonal manganites. However, the intensity profile from the as-grown domains shows that individual domain walls display a characteristic SEM signature, which extends over a length scale of about \SI{100}{\nano\metre} wide. We note that the latter is purely an SEM effect and can be explained based on the potential step between $+$P and $-P$ domains, as discussed in detail in ref. \cite{Nepijko2001}, and is not related to the actual width of the walls\cite{Holtz2017}. Thus, Figure \ref{fig2} (d) indicates that the domains are either fully poled or at least close to being fully poled. Although the electric fields generated by ion irradiation are difficult to quantify reliably, this observation leads us to the conclusion that they are comparable to the electric field required to achieve saturation polarization of the surface domains, which has been reported to be on the order of \SI{100}{\kilo\volt\per\centi\metre}. \cite{Kuerten2020}

Upon closer inspection and comparison of the as-grown and  poled domain structures, we find that the contracted domains form meandering lines that tend to follow the domain wall positions associated with the as-grown domain structure. This suggests that the domains do not shrink symmetrically, but rather tend to keep one domain wall at its original position. A possible explanation for this behavior is the interaction of the domain walls with intrinsic point defects, promoting domain wall pinning\cite{Smabraten2020}. Aside from the asymmetric domain switching, the SEM data shows a dark contrast around the contracted $+$P domains as briefly mentioned above. We propose that this contrast originates from the subsurface domain morphology as sketched in Figure \ref{fig2} (f). Analogous to the work by Kuerten et al.\cite{Kuerten2020}, it is likely that switching occurs only near the sample surface. As a consequence, the domain walls bent away from their ideal charge-neutral state form positively charged head-to-head walls, which could alter the electrostatic conditions near the surface and thereby the local electron yield. Another possibility is the switching revealing the unscreened polarization charge at the surface.

In order to avoid the possibility of ion-beam related domain wall pinning (as suggested by Figure \ref{fig2}), we also investigate reversible domain switching purely via electron irradiation. As illustrated in Figure \ref{fig0} (d), electron irradiation can lead to both negative and positive surface charging, depending on the incident energy $qV$. The corresponding switching experiment is shown in Figure \ref{fig3}. Figure \ref{fig3} (a) is obtained at \SI{1}{\kilo\volt}, showing the as-grown domain structure, analogous to Figure \ref{fig0} (b). In contrast to imaging at \SI{3}{\kilo\volt}, however, continuous imaging at lower voltage causes positive charging, leading to polarization switching as presented in Figure \ref{fig3} (b). Qualitatively, we observe the same behaviour as for ion beam-induced switching (Figure \ref{fig2}), that is, $+$P domains contracting to narrow lines at the surface. The SEM image shown in Figure \ref{fig3} (c) is also recorded at \SI{1}{\kilo\volt}, but immediately after repeated imaging at \SI{3}{\kilo\volt}, which neutralizes the positive charge that has previously built up. Comparison of Figures \ref{fig3} (a) and (c) show that by removing the surface charges, the original domain state can be fully restored. Close inspection of Figure \ref{fig3} (b) shows that in addition to the line-shaped domains, loop-shaped domains are created as shown in the inset. The formation of these loop-shaped domains cannot be explained by domain wall movement alone, demonstrating that polarization reversal also occurs via the nucleation and growth mechanism.

In summary, the switching experiments in Figures \ref{fig2} and \ref{fig3} demonstrate that both ion and electron beam irradiation can be used to switch the improper ferroelectric domains in \ce{ErMnO3}. By irradiating the sample with electrons with an energy $\approx qV_2$ (non-charging conditions), the as-grown domain state is recovered. The results are consistent with the switching behavior observed when using electrical contacts  \cite{Jungk2010, Han2013, Kuerten2020}, and can be explained assuming partial domain reversal at the surface and in surface-near regions as illustrated in Figure \ref{fig2} (f). 

\begin{figure}[H]
\includegraphics{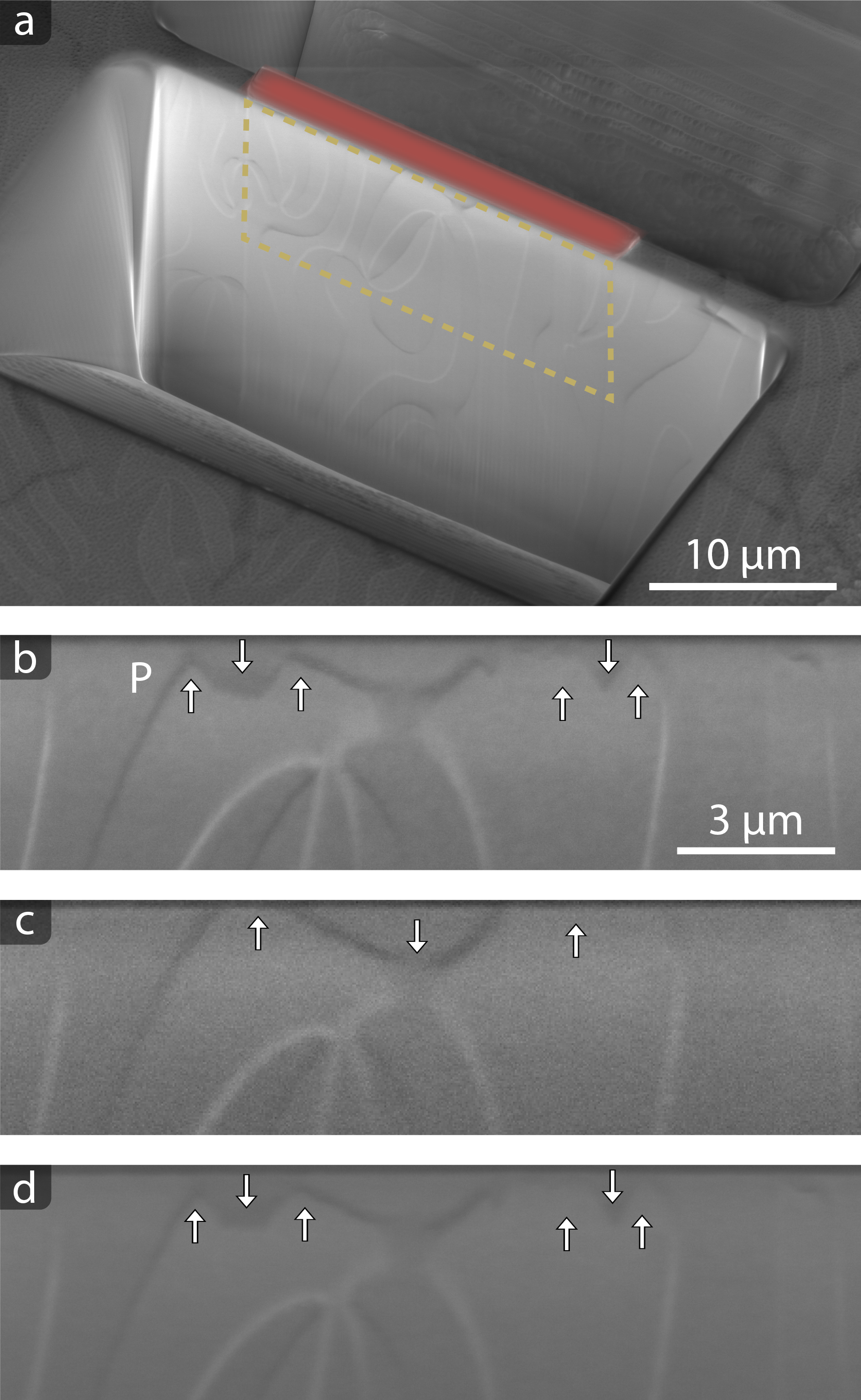}
 \caption{\label{fig1} Cross-sectional lamella cut from into a crystal with out-of-plane polarization. (a) Overview SEM image showing the isolated lamella. The face (in-plane polarization) is marked by the yellow rectangle, while the polar top surface is marked in red. (b) Lamella face after ion irradiation of top (polar) surface, showing that near the sample surface  $+$P domains contract and new $-$P domains arise. (c) After repeated electron beam scans of the lamella face, the as-grown domain configuration is recovered, with the domain walls returning to their original positions. (d) Repeated ion beam exposure of the top surface brings the domains back to the poled state in (b).}
 \end{figure}

To verify the hypothesis of partially switched domains near the surface and gain insight into the sub-surface switching behaviour, we prepare a cross-sectional lamella with in-plane polarization by FIB-milling of trenches in an out-of-plane polarized crystal, shown in Figure \ref{fig1} (a). Figure \ref{fig1} (b) shows the lamella face (marked by the yellow dashed rectangle in Figure \ref{fig1} (a))  imaged at \SI{2}{\kilo\volt}. The polarization directions are indicated by the arrows. In this orientation, no domain contrast is visible. Instead, we see only the domain walls with distinct contrast between the head-to-head and tail-to-tail domain walls. The dark walls are insulating head-to-head walls, while the bright are conductive tail-to-tail walls \cite{Mosberg2019}. The domain configuration in Figure \ref{fig1} (b) is observed after cutting out the lamella, i.e., after irradiating the top polar face (marked red in Figure \ref{fig1} (a)) with the \ce{Ga+} beam, inducing positive surface charging and contraction of $+$P domains as presented in Figure \ref{fig2}. Importantly, the cross-sectional SEM data shows that the domains have indeed only switched in the region near the surface, extending down to around \SI{1.4}{\micro\metre}. Under repeated SEM scanning of the lamella face at \SI{3}{\kilo\volt} (i.e. electron irradiation removing the positive surface charge), the $+P$ domains expand again, and a more balanced distribution of $+$P and $-$P domains in the top part of the lamella is restored (see Figure \ref{fig1} (c)). The partially poled surface state shown in Figure  \ref{fig1} (d) is reached again after irradiating the top of the lamella with \SI{30}{\kilo\electronvolt} \ce{Ga+}.  

\section{Summary and Outlook} 

Our results show that charged particle irradiation is a viable method for controlling improper ferroelectric domains in \ce{ErMnO3}. In particular, we find that charging the surface directly allows for bypassing barrier layer effects which arise when using metallic contacts and have prevented the domain control at room temperature in previous studies \cite{Ruff2018, Kuerten2020}. An open challenge associated with the irradiation approach as used here is, however, to adequately estimate the emergent electrical fields, which currently prohibits quantitative measurements.

 A back-of-the-envelope estimate made based on the imaging parameters used in Figure \ref{fig3} (beam current: \SI{0.1}{\nano\ampere}, pixel size: \SI{13.5}{\nano\metre}, dwell time: \SI{5}{\micro\second}) suggests that each scan enhances the surface charge density by about \SI{274}{\micro\coulomb\per\square\centi\metre}. This plane charge corresponds to an electric field of \SI{1.55e6}{\kilo\volt\per\centi\metre} in vacuum, which exceeds the coercive field $E_c$ of \ce{ErMnO3} ($E_C \approx \SI{30}{\kilo\volt\per\centi\metre}$)\cite{Han2013}. We note, however, that leakage currents, as well as electron emission and recapture drastically alter the charging conditions, and need to be controlled adequately in order to enable future quantitative measurements.
 
In conclusion, the combination of FIB and SEM enables contact-free manipulation of improper ferroelectric domains and analysis of their switching dynamics in 3D. We have demonstrated that polarization reversal in \ce{ErMnO3} occurs via both movement of domain walls and nucleation and growth of new domains, clarifying the switching behavior at the level of the domains. The possibility of {\em in-situ} nanostructuring with FIB allows for varying the boundary conditions, providing new opportunities for studying ferroelectric domain switching in confined geometries without the need for electrical contacts.

\begin{acknowledgments}
The authors acknowledge NTNU for support through the Enabling technologies: NTNU Nano program, the Onsager Fellowship Program and NTNU Stjerneprogrammet. The Research Council of Norway is acknowledged for financial support to the Norwegian Micro- and Nano-Fabrication Facility, NorFab, project number 245963/F50.  D.M. further acknowledges funding from the European Research Council (ERC) under the European Union’s Horizon 2020 research and innovation programme (Grant agreement No. 863691). Z.Y. and E.B. were supported by the U.S. Department of Energy, Office of Science, Basic Energy Sciences, Materials Sciences and Engineering Division under Contract No. DE-AC02-05-CH11231 within the Quantum Materials Program-KC2202.
\end{acknowledgments}

\section{Data Availability Statement}
The data that support the findings of this study are available from the corresponding author upon reasonable request.

\section{References}
\bibliography{bibliography.bib}

\end{document}